\documentclass[aps,pra,superscriptaddress,twocolumn,showpacs]{revtex4}
\usepackage{graphicx} 
\usepackage{amsmath}
 \usepackage{amssymb}

\newcommand{\beq}{\begin{equation}}
\newcommand{\enq}{\end{equation}}
\newcommand{\bea}{\begin{eqnarray}}
\newcommand{\ena}{\end{eqnarray}}

\newcommand{\ra}{\rangle}

\begin{document}

\title{Dynamical instability and loss of $p$-band bosons in optical lattices}
\author{J.-P. Martikainen}
\email{jpjmarti@nordita.org}
\affiliation{NORDITA, 106 91 Stockholm, Sweden}
\date{\today}

\begin{abstract}
We study how the bosonic atoms on the excited $p$-band of an optical lattice
are coupled to the lowest $s$-band and the $2$nd excited $d$-band.
We find that in some parameter regimes the atom-atom interactions 
can cause a dynamical instability of the $p$-band atoms towards 
decay to the $s$ and $d$-bands. Furthermore, 
even when dynamical instability is not expected
$s$- and $d$-bands can become substantially populated.
We also find that, the stability properties of the excited bands can be
improved in superlattices.
\end{abstract}

%\narrowtext
\maketitle 

\section{Introduction}
\label{sec:intro}
The study of cold atoms in optical lattices has seen a
dramatic experimental progress in the recent 
past~\cite{Bloch2008a,Lewenstein2007a}.
Due to realization of optical lattices, low densities, and low temperatures, 
a fantastic degree of control has been obtained which has made detailed studies 
of strongly correlated quantum systems possible. For example,
the Mott-superfluid transition~\cite{Jaksch1998a,Greiner2002a} 
has been successfully observed in optical lattices. 

The early experiments were confined to the lowest band and while 
increasing interactions can make excited band populations 
non-negligible~\cite{Kohl2006a}, the lowest band still dominates.  
Experimentally, however, atomic population residing on the excited bands can be 
obtained by coupling atoms from the lowest band to the excited bands. 
This has been experimentally demonstrated by accelerating the lattice for a short period~\cite{Browaeys2005a} or by coupling atoms from the lowest band Mott insulator into the first excited $p$-band of the lattice via Raman transitions between bands~\cite{Mueller2007a}. 
In this latter study, it was in particular found that the 
lifetimes of $p$-band atoms can be considerably 
longer than the tunneling time-scale in the lattice and it was 
also possible to 
explore how coherence of bosons on the excited band was established. 
Very recently $p$-band bosons and their orbital ordering
in the superfluid phase were studied experimentally in a bipartite optical 
lattice~\cite{Wirth2010a}. In this experiment atoms were transferred
to the $p$-orbitals by deforming the superlattice in such a way that
atoms originally in the lowest orbital of certain sites, could tunnel
to the $p$-orbital of the neighboring site.

The exciting possibilities in the physics with higher band atoms
has naturally attracted also theoretical attention. $P$-band bosons
have been studied in equilibrium~\cite{Isacsson2005a} and as a platform
to realize interesting correlated quantum 
states~\cite{Scarola2005a,Liu2006a,Xu2007a,Collin2010a}. Atoms
on the $p$-band have also been found to have 
interesting rotational properties~\cite{Umucalilar2008a}.
Furthermore, there are theoretical studies where multi-orbital description
of bosons is found necessary since it can induce important corrections
to the simplest lowest band Hubbard model~\cite{Sakmann2009a,Mering2010a}.

The purpose if this article is to point out the existence of dynamical
instability which can adversely affect the stability of the $p$-band atoms
in the superfluid phase. This instability is caused by the scattering
of $p$-band atoms into $s$- and $d$-band atoms. While the anharmonicities
in an optical lattice can make this process off-resonant for all
quasi-momenta, interplay between band structure and atom-atom interactions
can still induce instability. We also find that even in the absence
of dynamical instability, the $s$- and $d$-bands can become substantially
populated. Furthermore, dynamical instabities can be removed 
in superlattices~\cite{Wirth2010a}.

The paper is organized as follows. In Sec.~\ref{sec:theory} we outline a
theory which captures the essential physics of the loss of $p$-band
atoms to $s$ and $d$-bands
Furthermore, in this section
we discuss realistic magnitude of various parameters occurring in our theory.
 In Sec.~\ref{sec:dynamics} we derive and solve the 
Bogoliubov-de Gennes equations for the collective modes and find
the possibility of dynamical instability.  
In Sec.~\ref{sec:MI} we point out the importance
of condensation for the occurrence of dynamical instability, but
find that $s$- and $d$-bands can be substantially populated even
in the absence of dynamical instability.
In Sec.~\ref{sec:superlattices} we analyze
similar processes in superlattices and find that there
the dynamical instabilities can be removed.
We end with few concluding remarks in Sec.~\ref{sec:conclusions}.

\section{Formalism}
\label{sec:theory}
\begin{figure}
\includegraphics[width=\columnwidth]{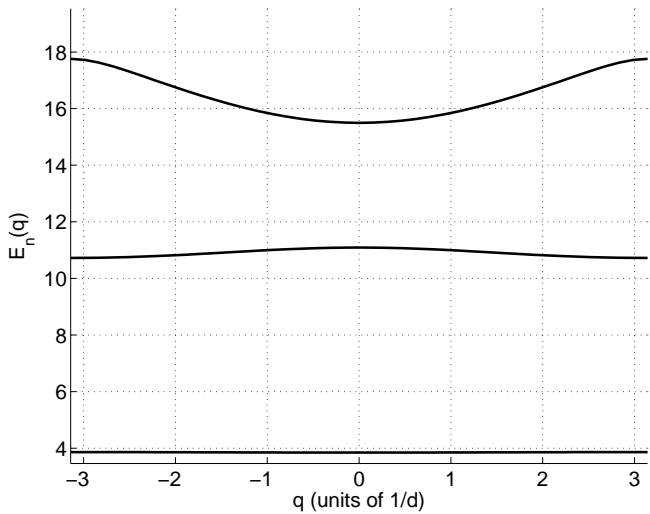}
\caption{A  band structure in units of $E_R$ for the three lowest bands
for a lattice depth of $V_L=17\, E_R$.}
 \label{fig1_bands}
\end{figure}

Let us consider bosonic atoms at zero temperature in an optical lattice
\beq
V({\bf r})=V_L \sin^2\left(\frac{\pi x}{d}\right)+
V_D\left[\sin^2\left(\frac{\pi y}{d}\right)
+\sin^2\left(\frac{\pi z}{d}\right)\right],\nonumber
\enq
where $d=\lambda/2$ is the lattice constant.
To simplify the formalism we assume a similar lattice potential
as in the experiment by M\"uller {\it et al.}~\cite{Mueller2007a}
so that the lattice is deep in $y$- and $z$-directions with depth 
$V_D=55\, E_R$, where $E_R=\hbar^2k^2/2m$ is the recoil energy, $m$ the atomic
mass, and the wavevector $k$ is related to the laser wavelength 
($\lambda=843\, {\rm nm}$)
through $k=2\pi/\lambda$. 
Throughout our calculation we will further assume
that we are dealing with $^{87}{\rm Rb}$ atoms. Since the lattice is deep
is two directions the dynamics in these directions is frozen and the
system is effectively one-dimensional. For the atomic wavefunctions along 
$y$- and $z$-directions we use the Gaussian ground states corresponding
to the harmonic oscillators along $y$ and $z$.
Using this effectively one-dimensional potential together with the usual 
kinetic energy operator ${\hat K}=-\hbar^2\nabla^2/2m$, we get an ideal Hamiltonian
$H_{lat}={\hat K}+V({\bf r})$ from which 
we numerically calculate the dispersions~\cite{Kittel} $E_s(q)$,$E_p(q)$, and
$E_d(q)$ for the  $3$-lowest ($s$, $p$, and $d$) bands as a function
of quasi-momentum $q$. Typical band structure is shown in Fig.~\ref{fig1_bands}.

Atoms interact through the two-body interaction~\cite{Pethick2001a}
$g/2\int d{\bf r} \hat\psi^\dagger({\bf r})\hat\psi^\dagger({\bf r})
\hat\psi({\bf r})\hat\psi({\bf r})$, where
$\hat\psi({\bf r})$ ($\hat\psi^\dagger({\bf r})$) is the bosonic 
annihilation (creation) operator and $g=4\pi\hbar^2 a/m$. 
For $^{87}{\rm Rb}$  atoms the scattering length $a\approx 110 a_0$.
Our main interest is to explore the role of the processes that transfer
atoms from the condensed $p$-band to the lowest 
$s$- and the second excited $d$-bands.
For this purpose we expand the field-operators in terms of the
field-operators $\hat{\psi}_\alpha$ for the three different bands.
\beq
\hat\psi({\bf r})=\phi_p({\bf x})\hat{\psi}_p+\phi_s({\bf x})\hat{\psi}_s
+\phi_d({\bf x})\hat{\psi}_d.
\enq
For the pure Bose-Einstein condensate in the $p$-band at quasi-momentum $q$,
$\phi_p({\bf x})$ corresponds to the Bloch-wavefunction $u_q^{p}({\bf x})$
at quasi-momentum $q$. The energy of the $p$-band atom is minimized
with the choice $q=\pi/d$. Likewise the functions $\phi_s({\bf x})$ and
$\phi_d({\bf x})$ are related to the Bloch-wavefunctions $u_q^{s,d}({\bf x})$
which dominate the dynamics at short times.
Substituting the above expansion, we find a generic interaction term relevant
to our discussion
\begin{eqnarray}
H_I&=&\frac{g_{pp}}{2}\hat\psi_p^\dagger\hat\psi_p^\dagger\psi_p\psi_p
+2n_p\left(g_{sp}n_s+g_{pd}n_d\right)\\
&+&2g_{ppsd}n_p\left(\hat\psi_s^\dagger\hat\psi_d+h.c\right)
+g_{ppsd}\left(\hat\psi_s^\dagger\hat\psi_d^\dagger\hat\psi_p\hat\psi_p+h.c\right)
,\nonumber
\label{eq:HI}
\end{eqnarray}
where $n_\alpha=\hat\psi_\alpha^\dagger\hat\psi_\alpha$.
We further assumed that $s$- and $d$-bands are almost
unpopulated by dropping terms which were of higher order in
$\hat{\psi}_s$ and $\hat{\psi}_d$ and dropped the far detuned processes
$2p\leftrightarrow 2s$ and $2p\leftrightarrow 2d$.
The form of the interaction Hamiltonian here is quite general and
only assumes the validity of the expansion into $3$ modes and the
weak population of $2$ of these modes.
The various interaction parameters are defined by the integrals
$g_{\alpha\beta}=\int d{\bf r} |\phi_\alpha({\bf x})|^2|\phi_\beta({\bf x})|^2$
and $g_{ppsd}=\int d{\bf r}\phi_p({\bf x})^2\phi_s({\bf x})^*\phi_d({\bf x})^*$.
It should be noted that in estimating the interaction strengths,
rather than using Bloch-wavefunctions
for $\phi_\alpha({\bf x})$, we could also have used the better localized
Wannier states, or their harmonic approximations without affecting
our results qualitatively. Even the quantitative differences are not dramatic
since the estimated interaction strengths are not very different
in the parameter regimes we consider.

The non-interacting part of the Hamiltonian is simply given by
$
H_0=E_sn_s+E_pn_p+E_dn_d
$
so that our total Hamiltonian is the sum $H=H_0+H_I$.
Note that here we also make an approximation that 
the atoms are sufficiently localized
so that tunneling dynamics plays only a minor role at the timescales
of interest. However, tunneling effects are to some extent still included
since they influence the band structure and are consequently incorporated
into the energy levels $E_n$. This approximation
is worst for the $d$-band (which is initially unoccupied)
whose bandwidth is by far the largest. However,
tunneling processes in the $d$-band would mainly give rise to (in practice)
irreversible 
atoms loss from the $p$-band since the more mobile $d$-band atoms
are free to relax towards their zero quasi-momentum minimum, where the process
$2p\leftrightarrow s+d$ is very far detuned. 

The detuning $\delta(q,k)=2E_p(q)-E_s(q-k)-E_d(q+k)$ which is related to the 
process $2p\leftrightarrow s+d$ described by the terms 
$\hat{\psi}_s^\dagger\hat{\psi}_d^\dagger\hat{\psi}_p\hat{\psi}_p+h.c.$ 
in the Hamiltonian, plays an important role for the stability
of the system. If one applies Fermi's golden rule to study
the life-time of $p$-band atoms~\cite{Isacsson2005a}, the lowest order
processes are no longer energetically allowed if the lattice is deeper
than about $V_L=18\,E_R$, since then $\delta(q,k)$ is non-vanishing
for all $q$ and $k$. If the process is always detuned 
for the $p$-band atoms at $q=\pi/d$ the detuning is minimized
with the choice $k=0$. Even if the detuning is zero somewhere, for the
lattice depths considered here, the resonance is still located close to $k=0$
so we choose $k=0$ without introducing large additional uncertainties when
computing the coupling coefficients of the interaction Hamiltonian.
For very deep lattices the detuning approaches a constant
value and in this sense the system differs from a harmonic oscillator
even in this limit~\cite{Collin2010a}. However, relative to the onsite
trap energy scale $\hbar\omega$ the detuning does approach zero
since $\omega\propto \sqrt{V_L}$.

\begin{figure}
\includegraphics[width=\columnwidth]{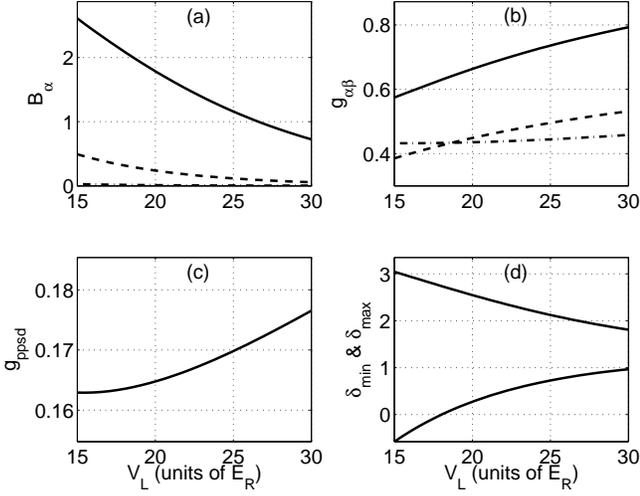}
\caption{Typical behavior for the relevant parameters as a function
of lattice depth $V_L$. In (a) we show the bandwidths 
(solid line for the $d$-band, dashed for the $p$-band, and dot-dashed
for the $s$-band). The bandwidth is defined as a difference between
the maximum energy within the band and the minimum energy within the band.
In (b) we show $g_{pp}$ (solid), $g_{sp}$ (dashed), and
$g_{pd}$ (dot-dashed), and (c) shows the behavior
of the coupling $g_{ppsd}$ responsible for population transfer away from
the $p$-band. Finally (d) shows the minimum and maximum values
of the detuning $\delta(q,k)$. When the minimum value of $\delta(q,k)>0$ 
the process $2p\leftrightarrow s+d$ is detuned for all quasi-momenta.
All energies in the $y$-axis are expressed in terms of the recoil energies $E_R$.}
 \label{fig2_parameters}
\end{figure}

In Fig.~\ref{fig2_parameters} we summarize the behavior of various
parameters of our model as the lattice depth is varied. It is to be noticed
that coupling coefficients appearing in $H_I$ are typically of the same
order of magnitude as the detuning $2E_p(q)-E_s(q)-E_d(q)$. For this reason
the system can be poorly described both in the narrow band as well as in
the wide band limit.

\section{Nonlinear condensate dynamics}
\label{sec:dynamics}
In the previous section we derived a minimal theory to describe
how $p$-band atoms are coupled to the $s$- and $d$-bands. Let us
now investigate what this model implies for the dynamics of pure
Bose-Einstein condensates. In this limit we can describe the
system in terms of complex amplitudes $\psi_s(t)$,  $\psi_p(t)$ ,
and $\psi_d(t)$ which are normalized to the number of particles
per site i.e. $|\psi_s|^2+|\psi_p|^2+|\psi_d|^2=n_T$.
This description also follows when the onsite wavefunctions are described
by coherent states and provides an accurate picture of the system
deep in the superfluid regime~\cite{Collin2010a}.
The equations of motions for the order parameters are given by
\beq
i\frac{\partial\psi_s}{\partial t}
=\left(E_s+2g_{sp}n_p\right)\psi_s+2g_{ppsd}n_p\psi_d
+g_{ppsd}\psi_p^2\psi_d^*,
\label{eq:psis}
\enq
\begin{eqnarray}
i\frac{\partial\psi_p}{\partial t}
&=&\left(E_p+g_{pp}n_p+2g_{sp}n_s+2g_{pd}n_d\right)\psi_p
\nonumber\\
&+&2g_{ppsd}(\psi_d\psi_s^*+c.c)\psi_p
+2g_{ppsd}\psi_p^*\psi_d\psi_s,
\label{eq:psip}
\end{eqnarray}
and
\beq
i\frac{\partial\psi_d}{\partial t}
=\left(E_d+2g_{pd}n_p\right)\psi_d+2g_{ppsd}n_p\psi_s
+g_{ppsd}\psi_p^2\psi_s^*.
\label{eq:psid}
\enq
In order to study the stability
of the system we now derive the relevant Bogoliubov-de Gennes equations.
Since the system is initially prepared in the $p$-band 
we have (at short times) 
$\psi_s(t)=\delta\psi_s e^{-iE_s't}$, $\psi_d(t)=\delta\psi_d e^{-iE_d't}$,
and $\psi_p(t)=\left[\sqrt{n_T}+\delta\psi_p\right]e^{-iE_p't}$,
where symbol $\delta$ is used to indicate a small quantity.
Furthermore, we defined $E_s'=E_s+2g_{sp}n_T$, $E_d'=E_d+2g_{pd}n_T$,
and $E_p'=E_p+2g_{pp}n_T$. By keeping only terms which are lowest order in 
$\delta\psi_\alpha$ we get a pair of equations
\beq
i\dot{\delta\psi_s}=2g_{ppsd}n_Te^{-i(E_d'-E_s')t}\delta\psi_d
+g_{ppsd}n_Te^{-i\delta't}\delta\psi_d^*\nonumber
\enq
and
\beq
i\dot{\delta\psi_d}=2g_{ppsd}n_Te^{+i(E_d'-E_s')t}\delta\psi_s+g_{ppsd}n_Te^{-i\delta't}\delta\psi_s^*,\nonumber
\enq
where $\delta'=2E_p'-E_s'-E_d'$. The first terms on the right of these
equations can be dropped since they are oscillating rapidly and 
their time average vanishes over the timescales of interest.
By defining
\beq
\delta\psi_s(s)=\left[u_se^{-i\omega t}+v_s^*e^{+i\omega t}\right]e^{-i\delta'/2t}
\enq
and
\beq
\delta\psi_d(s)=\left[u_de^{-i\omega t}+v_d^*e^{+i\omega t}\right]e^{-i\delta'/2t}
\enq
we find the Bogoliubov-de Gennes eigenvalue problem for collective modes
\beq
\omega\hat{\eta}{\bf \lambda}=\hat{M}{\bf \lambda}
\enq
described by their frequency $\omega$ as well as amplitudes $u_\alpha$
and $v_\alpha$. Here ${\bf \lambda}^T=\left(u_s, v_s,u_d,v_d\right)$,
\beq
\hat{\eta}=\left(\begin{array}{cccc}
1 &0 &0 &0\\
0 &-1 &0 &0\\
0 &0 &1 &0\\
0 &0 &0 &-1\\
\end{array}\right),
\enq
and
\beq
\hat{M}=\left(\begin{array}{cccc}
-\delta'/2 &0 &0 &g_{ppsd}n_T\\
0 &-\delta'/2 &g_{ppsd}n_T &0\\
0 &g_{ppsd}n_T &-\delta'/2 &0\\
g_{ppsd}n_T &0 &0 &-\delta'/2\\
\end{array}\right).
\enq
This eigenvalue problem has solutions
\beq
\omega=\pm\frac{1}{2}\sqrt{\delta'^2-4(g_{ppsd}n_T)^2}.
\label{eq:BdGmodes}
\enq
Importantly, these solutions are imaginary and indicate dynamical instability
if $|\delta'|<2g_{ppsd}n_T$. In a parameter regime where dynamical instability
exists a system initially prepared on the $p$-band will
lose atoms to the $s$- and $d$-bands at a rate given by
\beq
\Gamma=\sqrt{|\delta'^2-4(g_{ppsd}n_T)^2|}.
\enq
It is also easy to see when this instability is more likely to occur.
For relatively deep lattices $\delta=2E_p-E_s-E_d>0$ since the
anharmonicities of the lattice potential shifts the $d$-states lower most
relative to the harmonic oscillator energy levels.
On the other hand  the effective detuning is given by
 $\delta'=\delta+2n_T(g_{pp}-g_{sp}-g_{pd})$ and will move closer
to zero if $g_{pp}-g_{sp}-g_{pd}<0$, a condition which usually holds since
all the coupling coefficients have similar magnitudes. Then whenever
$\delta'$ becomes small relative to $g_{ppsd}n_T$ 
(which increases with lattice depth)
dynamical instabilities can be expected.

\begin{figure}
\includegraphics[width=\columnwidth]{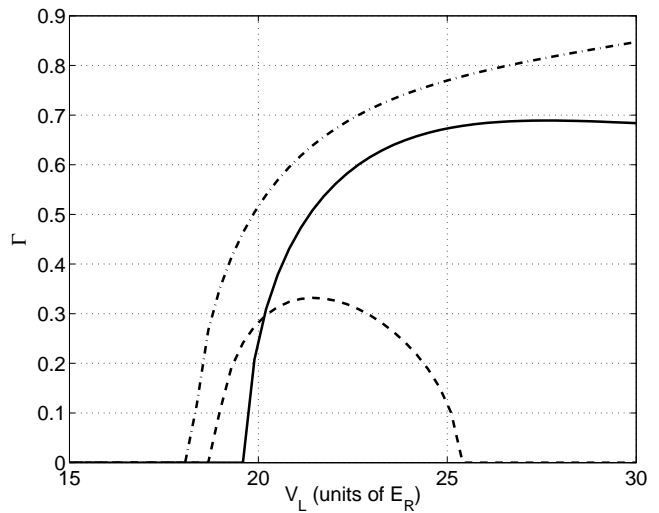}
\caption{Loss rate $\Gamma$ of the $p$-band atoms as a function
of the lattice depth $V_L$ for $n_T=1$ (dashed) and $n_T=2$ atom per site.
For comparison, with a dot-dashed line, we show the result where
onsite wavefunctions were assumed to be harmonic oscillator wavefunctions
and $n_T=2$.
The loss rate is given in units of $1/\tau=E_R/\hbar$. For
$^{87}{\rm Rb}$ atoms $\tau$ is about $49\, {\rm \mu s}$.}
 \label{fig3_lossrate}
\end{figure}

In a Fig.~\ref{fig3_lossrate} we show the loss rate of the $p$-band atoms
as a function of lattice depth for two different onsite atom numbers.
This figure suggests that for small onsite atom numbers
a dynamical instability can be present 
with realistic trap parameters above $V_L\approx 19\, E_R$
and below $V_L=25\, E_R$
at which the system is expected to be in the superfluid 
regime~\cite{Mueller2007a}.  For higher atom numbers the region of 
instability is greatly increased and extends into the $p$-band
Mott-insulating region where our assumption of coherent states
is clearly invalid.
Furthermore, we find that the timescale for the instability 
is substantially less
than the timescale for the tunneling of the $p$-band atoms justifying
our earlier approximation to ignore tunneling dynamics.
Interestingly, the dynamical instability is expected for lattice
depths at which the lowest order Fermi's golden rule becomes
inapplicable~\cite{Isacsson2005a}.

\begin{figure}
\includegraphics[width=\columnwidth]{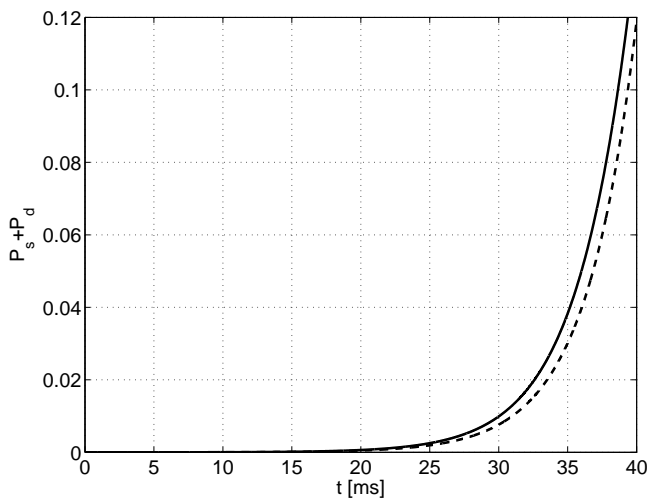}
\caption{Comparison between the Bogoliubov- de Gennes result (dashed line) 
for the s- and d-band populations as a function of time
and the solution to the full Gross-Pitaevskii equations (solid line). 
The lattice depth
was chosen as $V_L=20\, E_R$ and $n_T=1$ so that the system was predicted to be
in the unstable regime. Initial state had a $10^{-6}$ population
on s- and d-bands (to seed the dynamics) and the phases of the amplitudes
were taken to be zero.}
 \label{fig4_comparison}
\end{figure}

We have also solved the Eqs. (\ref{eq:psis})-(\ref{eq:psid})
numerically even by modifying them to include interactions between
$s$- and $d$-band atoms. We compare the
Bogoliubov- de Gennes approach with the full Gross-Pitaevskii
equations in Fig.~\ref{fig4_comparison}.
As is clear the agreement between the numerical
and analytical Bogoliubov- de Gennes 
solution is very good when the p-band population
is dominant and the small difference is mainly
due to the facts that short time dynamics has some sensitivity to
phases of the amplitudes and that numerics 
also included many far off-resonant processes
giving rise to fast oscillations on a much shorter timescale. These processes
were ignored in the Bogoliubov- de Gennes approach derived earlier.
If in the dynamically unstable region
the initial state has a substantial $s$-band population the time-evolution
becomes more sinusoidal, but even then the maximum $d$-band population
is substantial and cannot be ignored.

The numerical solutions also reveal a revival in the $p$-band population
after the initial exponential loss. However, we believe that this revival
is unlikely to persist in a realistic system, due to the high mobility
of the $d$-band atoms. It is more likely that those atoms coupled
to the $d$-band will be irreversibly lost from the $p$-band.
Also, in a trap the density distribution of the more mobile
$d$-band atoms will be much broader since the length scale
for the density distribution in a harmonic trap is proportional to the
$B_d^{1/4}$, where $B_d$ is the $d$-band bandwidth. For this reason, many $d$-band atoms
end up spatially separated from the $s$- and $p$-band atoms residing
closer to the trap minimum.

\subsection{Role of number fluctuations}
\label{sec:MI}
\begin{figure}
\includegraphics[width=\columnwidth]{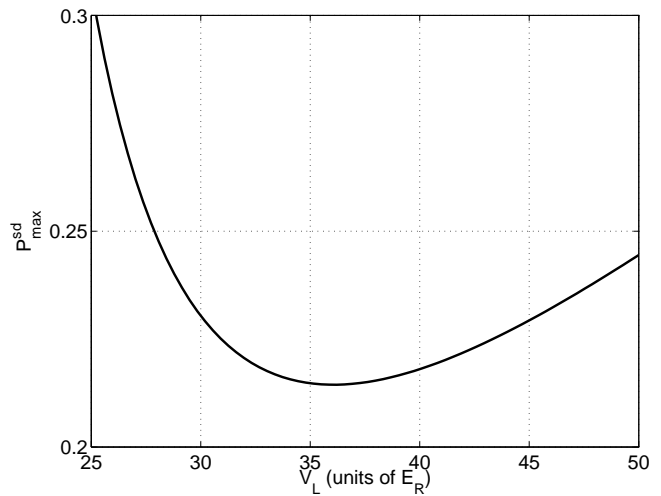}
\caption{The maximum population of the $|1,0,1\ra$-state when the 
number of atoms $n_T=2$ as a function of lattice depth.
The interaction parameters were estimated by approximating the 
lattice site by a harmonic oscillator while the energy levels were
computed with a band-structure calculation with a quasi-momentum $q=\pi/d$.
}
\label{fig5_nT2}
\end{figure}

In the previous section we described the system by assuming pure condensates
in all bands and found a possibility of dynamical instability. The question
then arises that what role did the assumption of condensation
actually play in our result? Alternatively, deep in the Mott-insulating regime
with $n_T=2$ atoms per site we can at short times use an ansatz 
\beq
|\psi(t)\ra=\psi_{sd}(t)|1,0,1\ra+\psi_{p}(t)|0,2,0\ra,
\enq
where $|n_s,n_p,n_d\ra$ is the Fock state. To the lowest order
in tunneling the many-body wavefunction factorizes into single site
solutions, so the solution in a single site amounts to a solutions
throughout the insulator.
Due to the presence
of the $|1,0,1\ra$-state we must furthermore include the 
density-density interaction $2g_{sd}n_sn_d$
between $s$- and $d$-bands into our
model. In the earlier mean-field theory this term was a small
quantity which could be safely ignored.
We then get (at short times)
\begin{eqnarray}
H|\psi(t)\ra&=&\left[(E_s+E_d+2g_{sd})\psi_{sd}+g_{ppsd}\sqrt{2}\psi_p\right]|1,0,1\ra
\nonumber\\
&+&\left[(2E_p+g_{pp})\psi_{p}+g_{ppsd}\sqrt{2}\psi_{sd}\right]|0,2,0\ra.
\end{eqnarray}
In this case 
we can easily solve the equations of motion for $\psi_p(t)$ and $\psi_{sd}(t)$ 
and find a simple Rabi-problem with purely real eigenenergies
\begin{eqnarray}
E_\pm(n_T=2)&=&\frac{2E_p+E_s+E_d+g_{pp}+2g_{sd}}{2}
\\
&&\hspace{-1cm}\pm\frac{1}{2}\sqrt{\left(2E_p-E_s-E_d+g_{pp}-2g_{sd}\right)^2+8g_{ppsd}^2}\nonumber
\end{eqnarray}
and an initial state $|\psi(t=0)\ra=|0,2,0\ra$ evolves in such a way that
the maximum population of the $|1,0,1\ra$ state is given by
\beq
P^{sd}_{max}=\frac{8g_{ppsd}^2}{(2E_p-E_s-E_d+g_{pp}-2g_{sd})^2+8g_{ppsd}^2}.
\enq
In Fig.~\ref{fig5_nT2} we show the maximum population of the 
$|1,0,1\ra$ state which gives in indication of how reliable a pure $p$-band
theory can be. Since our ansatz with a fixed atom number 
is expected to be reasonable only for deep lattices,
in this figure we estimated the interaction parameters
by approximating the lattice site with a harmonic oscillator. As is clear
from this figure,
substantial fraction of the $^{87}{\rm Rb}$
population can reside outside the $p$-band. In fact, 
$P^{sd}_{max}$ can even reach unity when $2E_p-E_s-E_d+g_{pp}-2g_{sd}=0$ 
which formally happens somewhat below the lattice depth of $V_L=20\, E_R$.
However, in this regime stronger tunneling invalidates
the simple three-state description as well as the assumption of fixed 
onsite atom number.

With exactly $n_T=1$ atoms per site interactions do not contribute
and instability is not expected.
These simple exercises indicate an important role of onsite
number fluctuations in feeding the dynamical instability. What happens
if the system is neither a Mott-insulator nor a pure condensate,
is unclear. Presumably, the system is more stable than our calculations
assuming a pure condensate
suggest, but the stability cannot be taken for granted.

\subsection{Dynamical instabilities in superlattices}
\label{sec:superlattices}
Since $p$-band bosons have also been studied
in a bipartite optical lattice~\cite{Wirth2010a,Stojanovic2008a}, 
let us briefly discuss
the stability properties of $p$-band atoms in superlattices.
We will again assume the same deep optical lattice along $y$- and 
$z$ directions, but for the lattice along $x$-direction we
use a potential~\cite{Trotzky2010a}
\beq
V(x)=V_S\sin^2\left[\frac{\pi x}{d}\right]
-V_L\sin^2\left[\left(\frac{\pi}{2d}\right)(x+d)\right] 
\enq
which is characterized by the depths $V_S$ and $V_L$ of the short
and long lattices respectively. 

\begin{figure}
\includegraphics[width=\columnwidth]{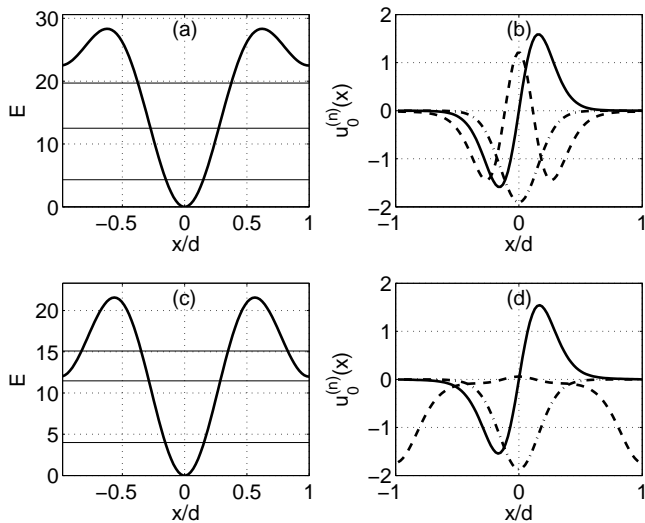}
\caption{Energy levels in units of $E_R$ (together with the lattice potential) 
and wavefunctions in a superlattice. We fixed the short lattice depth to
$V_S=15\, E_R$ while $V_L=1.5V_S$ in (a) and (b) or
$V_L=0.8V_S$ in (c) and (d). In (b) and (d) the dot-dashed line
is the wavefunction for the $s$-orbital, solid line for the $p$-orbital,
and the dashed line for the $d$-orbital.
}
 \label{fig6_superlattice}
\end{figure}

\begin{figure}
\includegraphics[width=\columnwidth]{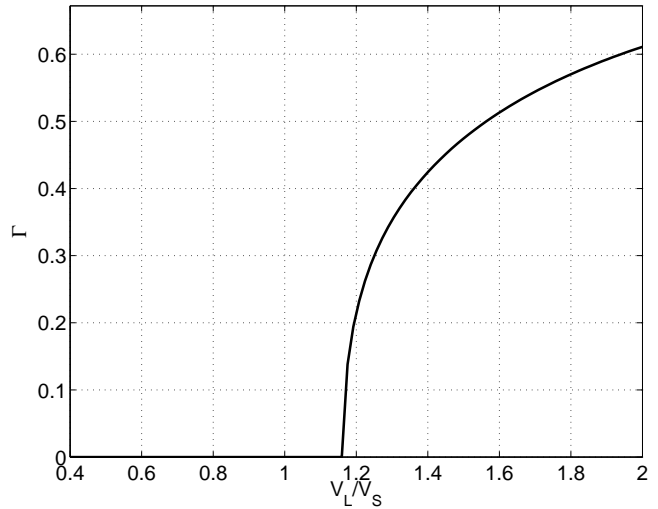}
\caption{Loss rate of the $p$-band atoms in a superlattice due to the dynamical
instability as a function of $V_L/V_S$ when the short lattice depth
was fixed to $V_S=15\, E_R$. The coupling coefficients and band energies
were estimated from the $q=0$ eigenstates and
the loss rate is given in units of $1/\tau=E_R/\hbar$.
}
 \label{fig7_superlatticeloss}
\end{figure}

In Fig.~\ref{fig6_superlattice}
we show examples of the energy levels together with the three lowest
eigenstates of this potential while in  Fig.~\ref{fig7_superlatticeloss}
we show an estimate of when dynamical instabilities might be expected.
As is clear from these figures  dynamical instabilities
might occur only when $V_L/V_S$ is fairly large and 
the eigenstates are well localized to the deep sites.
When $V_L/V_S$ becomes smaller instability is quickly suppressed.
The reason for this is twofold. First, as  $V_L/V_S$ becomes smaller
the energy levels become very different from
the energy levels of a harmonic oscillator at deep sites and this
implies an increase in the detuning $2E_p-E_s-E_d$ which reduces
the likelihood of dynamical instabilities. Second, while for large
$V_L/V_S$ all the wavefunctions are well localized to deep sites and their
overlaps are large, for smaller $V_L/V_S$ the $d$-orbital becomes
peaked in the shallow sites. When this occurs, the overlap between the
$d$-orbital and the $s$- and $p$-orbitals is drastically reduced
and $g_{ppsd}$ is reduced by more than an order of magnitude.
From this we can conclude that superlattices can enhance the stability
properties of $p$-band bosons. However, this happens at the cost
of greater hybridization of the $p$-orbitals in deep sites
with the $s$-orbitals in the shallow sites. In fact this hybridization
played an important role also in the experiment 
by Wirth {\it et al.}~\cite{Wirth2010a}.

\section{Conclusions}
\label{sec:conclusions}
We have pointed out a dynamical instability which can affect the
stability of $p$-band bosons in the broken symmetry phase.
All our numerical examples assumed $^{87}{\rm Rb}$ atoms and consequently
some details are expected to be different for different atoms.
In particular, the scattering length will be different for different
atoms or it can be tuned with Feshbach resonance and this can have a dramatic
effect on how big a role the coupling to the $s$- and $d$-bands plays. 
For convenience we restricted our discussions to an effectively
one-dimensional system. If the dimensionality is increased
we have several degenerate flavors in the $p$- and $d$-bands
and many more interaction channels. However, the 
Gross-Pitaevskii equations analogous to 
Eqs. (\ref{eq:psis})-(\ref{eq:psid}) would still look very similar
and the stability analysis is likely to reveal  similar 
instabilities as we found here for one-dimensional systems.

We found that while $p$-band leaks into $s$- and $d$-bands even in
the Mott-insulating regime,
for the dynamical instabilities to be present it was important
that the system is not a Mott-insulator. However,  even
in a Mott-insulating regime for $p$-band atoms, these
atoms are coupled to $s$ and $d$-bands.
Since the bandwidth for the $d$-band atoms is dramatically larger
than for lower bands, $d$-band atoms would typically be deep in the superfluid
regime. Under such circumstances the $p$-band atoms are coupled
to a coherent $d$-band and this coupling will induce number
fluctuations also on the $p$-band. How this effect affects 
the superfluid Mott-insulator transition
for $p$-band atoms has not yet been explored.

Finally it should be noted that the stability properties 
of fermionic atoms are likely to be better than those of bosonic atoms.
This is because with fermions one can populate the $p$-band by first
filling the lowest band. When the lowest band is filled, the
coupling between $p$, $s$, and $d$-bands becomes Pauli blocked.

%\bibliographystyle{apsrev}
%\bibliography{./bibli}

\end{document}